\documentclass{PoS}

\usepackage{bm}

\title{Numerical Tests of the Improved Fermilab Action}

\ShortTitle{Numerical Tests of the Improved Fermilab Action}

\author{C. DeTar\\
        Physics Department, University of Utah, Salt Lake City, Utah, USA\\
        E-mail: \email{detar@physics.utah.edu}}

\author{A. S. Kronfeld\\
        Theoretical Physics Department, 
        Fermi National Accerelator Laboratory,
        Batavia, Illinois, USA\\
        E-mail: \email{ask@fnal.gov}}

\author{\speaker{M. B. Oktay}\\
        Physics Department, University of Utah, Salt Lake City, Utah, USA\\
        E-mail: \email{oktay@physics.utah.edu}}

\abstract{Recently, the Fermilab heavy-quark action was extended to 
	  include dimension-six and -seven operators in order to reduce 
	  the discretization errors. In this talk, we present results
	  of the first numerical simulations with this action (the OK action),
          where we study the masses of the quarkonium and heavy-light 
	  systems. We calculate combinations of masses designed to test 
          improvement and
          compare results obtained with the OK action to their counterparts 
          obtained with the clover action.
          Our preliminary results show a clear improvement.}

\FullConference{The XXVIII International Symposium on Lattice Field Theory, 
		Lattice2010\\
		June 14-19, 2010\\
		Villasimius, Italy}

\begin{document}

\section{Introduction}

Simulating heavy quarks in lattice QCD 
is a challenging problem, because the quark mass~$m_Q$ and the 
accessible ultraviolet cutoff $a^{-1}$ are comparable.
Special care is needed to handle discretization errors 
\cite{Kronfeld:2003sd}.
In order to make accurate and reliable calculations of many Standard Model
parameters involving heavy quarks, one needs not only computer power
but also methodological improvements of the quark actions used. 
One line of attack is the Fermilab method \cite{El-Khadra:1996mp}, 
which starts with the clover action \cite{Sheikholeslami:1985ij} for 
Wilson fermions \cite{Wilson:1975hf}.
In the original work, interactions through dimension five were 
considered.
More recently, we extended the Fermilab method to an action (the OK 
action) with dimension-six and -seven 
interactions~\cite{Oktay:2008ex,Oktay:2002mj}.

Using power counting as a guide, Ref.~\cite{Oktay:2008ex} estimated 
that the OK action should reduce discretization effects for heavy 
quarks to $\sim1\%$ on, say, the MILC asqtad 
ensembles~\cite{Bazavov:2009bb}.
In this paper, we present the first numerical results obtained with the 
OK action, to test whether the theoretical improvement is realized in 
practice.
We compute combinations of rest masses and kinetic masses designed to 
test the improvement, without the need to tune the input parameters.

In Sec.~\ref{sec:action}, we briefly discuss and present 
the OK action. Section~\ref{sec:simulation} contains
the details of the simulations and our preliminary results for an
inconsistency $I$~\cite{CollinsYG}, 
and for the hyperfine splittings between rest
and kinetic masses for the heavy-heavy and heavy-light systems.
We discuss our results and future plans in section~\ref{sec:outlook}.

\section{OK Action and Tadpole Improvement}
\label{sec:action}

In this section, we briefly describe the OK~action, including the 
tadpole improvement~\cite{Lepage:1992xa} used in the simulation.
In general, the
Fermilab formalism 
calls for separate couplings for
spatial and temporal interactions. Fermilab actions have a smooth transition
to the $m_0a\to0$ and $m_0a\to\infty$ limits, but the short-%
distance coefficients depend on $m_0a$, $\zeta$ and the spatial Wilson
parameter $r_s$ in a non-trivial way \cite{El-Khadra:1996mp}. This
action reduces the discretization errors to 
$\mathrm{O}(a^2\Lambda^2b(m_Qa))$, where $b$ is a function that is 
bounded for all $m_Qa$. The OK~action includes
higher dimensional operators to further reduce the lattice spacing errors.

Reference~\cite{Oktay:2008ex} starts by considering all the operators of dimension 
six and seven with two effective field theories in mind: heavy-quark
effective theory (HQET) and the nonrelativistic QCD (NRQCD), 
appropriate to heavy-light and heavy-heavy systems, respectively.
The interactions are classified in powers of $\lambda$
(HQET, $\lambda\sim \Lambda/m_Q$, $\Lambda a$) or $v$
(NRQCD, relative internal velocity).
Once all the 
independent operators are identified, the redundant ones are eliminated
by means of field transformations, and the remaining 
couplings $c_i$ are determined via tree-level matching. 
Together with the one-loop matching of the
dimension five chromomagnetic interaction, this action is expected to
bring the discretization errors below the one-percent level \cite{Oktay:2008ex}. 

For coding, and especially for tadpole improvement of the couplings, it 
is convenient to write the OK action in the hopping-parameter form:
\begin{eqnarray}
S&=& \sum_x\bar{\psi}_x\psi_x
                  -\kappa_t\sum_x\bar{\psi}_x(1-\gamma_4)T_4\psi_x
                  -\kappa_t\sum_x\bar{\psi}_x(1+\gamma_4)T_{-4}\psi_x
                  \nonumber \\
 &-& \kappa_t\sum_{x,i}\bar{\psi}_x[(r_s\zeta+8c_4)
 -\gamma_i(\zeta-2c_1-12c_2)]T_i\psi_x \nonumber \\
 &-& \kappa_t\sum_{x,i}\bar{\psi}_x[(r_s\zeta+8c_4)
 +\gamma_i(\zeta-2c_1-12c_2)]T_{-i}\psi_x \nonumber \\ 
 &+& 
 \kappa_t\sum_x\bar{\psi}_x[2c_4+\gamma_i(c_1+2c_2)]T_i^2\psi_x 
 - \kappa_t(c_B\zeta+16c_5)\sum_x\bar{\psi}_xi\bm{\Sigma}\cdot\bm{B}_{\rm lat}\psi_x
 \nonumber \\              
 &+& 
 \kappa_t\sum_x\bar{\psi}_x[2c_4-\gamma_i(c_1+2c_2)]T_{-i}^2\psi_x
 -\kappa_tc_E\zeta\sum_x\bar{\psi}_xi\bm{\alpha}\cdot\bm{E}_{\rm lat}\psi_x 
  \nonumber \\
  &+&
 \kappa_tc_2\sum_{x,i\neq j}\bar{\psi}_x\gamma_i\{T_i\!-\!T_{-i},T_j\!-\!T_{-j}\}
  \psi_x  + 2\kappa_tc_{5}
  \sum_x\bar{\psi}_x\sum_i\sum_{i\neq j}
  \{i\Sigma_iB_{i\rm{lat}},(T_j\!+\!T_{-j})\}\psi_x \nonumber   \\
  &+&
  2\kappa_tc_3\sum_x\bar{\psi}_x\{\bm{\gamma}\cdot\bm{D}_{\rm lat},
  i\bm{\Sigma}\cdot\bm{B}_{\rm lat}\}\psi_x + 2\kappa_tc_{EE}\sum_x
  \bar{\psi}_x\{\gamma_4D_{4\rm{lat}},\bm{\alpha}\cdot\bm{E}_{\rm lat}\}\psi_x \nonumber
\end{eqnarray}
where $T_{\pm\mu}\psi(x) = U_{\pm\mu}(x)\psi(x\pm a\hat{\mu})$ with
      $U_{-\mu}(x) = U^\dagger_\mu(x-a\hat{\mu})$
and 
\begin{eqnarray}
    m_0a = \frac{1}{2\kappa_t}-(1+3r_s\zeta+18c_4). 
\end{eqnarray}
For further details and the matching conditions for the 
couplings~$c_i$, we refer the reader to Ref.~\cite{Oktay:2008ex}.

Given the large one-loop corrections that can arise in lattice 
perturbation theory~\cite{Lepage:1992xa}, before using the OK action in 
a numerical simulation we apply tadpole improvement to 
the matched couplings.
To carry out the tadpole improvement, it is convenient to write
$T_{\pm\mu} =  u_0[T_{\pm\mu}/u_0] = u_0\tilde{T}_{\pm\mu}$ where 
$u_0$ factors are absorbed into the couplings~$\tilde{c}_i$. In this way, 
one finds the relations between bare and tadpole improved coefficients 
\begin{eqnarray}
\tilde{\kappa}_t &=&  u_0\kappa_t, \\
\tilde{r}_s\tilde{\zeta} + 8\tilde{c}_4  &=& r_s\zeta+8c_4  \\
\tilde{\zeta}-2\tilde{c}_1-12\tilde{c}_2 &=& \zeta-2c_1-12c_2  \\
\tilde{c}_4 & = & u_0c_4, \\
\tilde{c}_1 + 2\tilde{c}_2 &=& u_0(c_1 + 2c_2)  \\ 
\tilde{c}_2 & = & u_0c_2, \\
\tilde{c}_B\tilde{\zeta} + 16\tilde{c}_5 & = & u_0^3(c_B\zeta+16c_5)  \\
\tilde{c}_E & = & u_0^3c_E, \\
\tilde{c}_3 & = & u_0^4c_3, \\
\tilde{c}_{EE} & = & u_0^4c_{EE},
\end{eqnarray}
where the last two follow because every term in the anticommutators has 
five links, while one power of $u_0$ is absorbed, as usual, into 
$\tilde{\kappa}_t$.
The matching conditions of Ref.~\cite{Oktay:2008ex} are then used, 
substituting 
\begin{equation}
\tilde{m}_0a = \frac{1}{2\tilde{\kappa}_t}-(1+3\tilde{r}_s\tilde{\zeta}
+18 \tilde{c}_4)     
\end{equation}
for $m_0a$.
Another condition for $\tilde{c}_5$ is needed, but it is not 
simple to express.
In the expansion of $B_{i{\rm lat}}$ in terms of $T$s, one finds that 
$\bar{\psi}_x\{i\Sigma_i B_{i{\rm lat}}, (T_j+T_{-j}) \} \psi_x$,
$j\neq i$,
has terms with both 3 and 5 links.
The 3-link terms arise from $T_{\pm\mu}T_{\mp\mu}=1$.
In the $c_3$ and $c_{EE}$ interactions, each 3-link term appears
twice, but with opposite sign, while here they have the same sign.
Coding this operator with the correct $u_0$ factors is currently underway
with USQCD software \cite{usqcd}.

\section{Simulations and Tests}
\label{sec:simulation}

We performed simulations on a ``medium coarse'' ($a\approx 0.15$~fm) $16^3\times48$
lattice with 2+1 flavors of sea quarks, $(am_l,am_s)=(0.029,0.0484)$.
From Ref.~\cite{Burch:2009az}, we had data available with the clover 
action, and here we used the OK action with similar statistics,
500 configurations with 4 time sources per configuration.
For the results presented here, we tadpole-improved the 
$\bar{\psi}_x\{i\Sigma_i B_{i{\rm lat}}, (T_j+T_{-j}) \} \psi_x$,
$j\neq i$, interaction with $\tilde{c}_5=u_0^4c_5$, pending completion 
of the code with proper tadpole improvement.
We also choose $\tilde{r_s}=\tilde{\zeta}
=\tilde{c}_B = 1$. This choice fixes the value of $\tilde{c}_4$ and 
$\tilde{c}_5$ while the rest of the coefficients also depend on the 
choice of $\tilde{\kappa}_t$. 

Since the action is designed to improve $O(p^4)$ terms, we need
to find observables to test these improvements. One such quantity is
a combination of masses introduced in 
Ref.~\cite{CollinsYG} and later discussed in Ref.~\cite{KronfeldUY}. Writing
the rest and kinetic meson masses $M_{1\bar{Q}q}$ and $M_{2\bar{Qq}}$ as
\begin{eqnarray}
M_{1\bar{Q}q} &=& m_{1\bar{Q}}+m_{1q} + B_{1\bar{Q}q}, \\
M_{2\bar{Q}q} &=& m_{2\bar{Q}}+m_{2q} + B_{2\bar{Q}q}, 
\end{eqnarray}
where the $m$s are quark masses and the $B$s are binding energies,
Ref.~\cite{CollinsYG} introduced the ``inconsistency combination''
\begin{equation}
I := \frac{2\delta M_{\bar{Q}q}-(\delta M_{\bar{Q}Q} + \delta M_{\bar{q}q})}
            {2M_{2\bar{Q}q}} =
    \frac{2\delta B_{\bar{Q}q}-(\delta B_{\bar{Q}Q} + \delta B_{\bar{q}q})}
            {2M_{2\bar{Q}q}}
\end{equation}
where $\delta M = M_2-M_1$, $\delta B=B_2-B_1$.
The rightmost expression follows from the definitions.

The binding energies $B_2$ stem from the $p^4$ terms in the 
action~\cite{KronfeldUY}.
By design, the the OK~action improves these terms, compared with the 
clover action.
Ideally, $\delta B$s and, hence, $I$ should vanish, and
we expect $I$ to be smaller with an improved action.
In order to compare the OK~action
with the clover action, we 
compute the rest and kinetic masses of heavy-light
and heavy-heavy systems at four different 
hopping-parameter values, $\tilde{\kappa}_t=0.042,0.040,0.038,0.036$.
Table~\ref{tab:mass} lists the obtained pseudoscalar kinetic masses,
to show the range of physical mass covered here.
\begin{table}[tbp]
\caption{Approximate values of the kinetic masses for the heavy-light and 
heavy-heavy systems obtained with the OK~action.
Thus, 0.042 is in the charm region and 0.036 is not yet in the bottom 
region.}
\label{tab:mass}
\begin{center}
\begin{tabular}{ccc}
\hline\hline
$\tilde{\kappa}_t$ & $M_2^{\mathrm{PS}}$ (Heavy-Light) [MeV]
         & $M_2^{\mathrm{PS}}$ (Heavy-Heavy) [MeV] \\
\hline
0.036 & 4418 & 7332 \\
0.038 & 3500 & 5778 \\
0.040 & 2680 & 4227 \\
0.042 & 1876 & 2736 \\
\hline\hline
\end{tabular}
\end{center}
\end{table}

\begin{figure}[btp]
\begin{center}
\includegraphics[width=.6\textwidth]{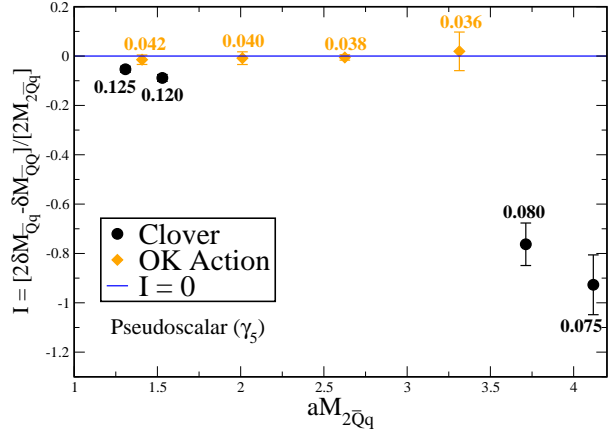}
\caption{Comparison of the inconsistency $I$ between the OK~action and 
the clover action. 
For the clover action 
$\kappa=0.122$ and $\kappa=0.076$ correspond to tuned $\eta_c$ and $\eta_b$
masses.}
\label{fig:fig1}
\end{center}
\end{figure}

Figure~\ref{fig:fig1} shows our results for the inconsistency $I$, 
together with results from an earlier study with the clover 
action~\cite{Burch:2009az}.
For simplicity, and without serious loss in the strength of the test, 
we omit the light-light mass difference $\delta M_{\bar{q}q}$.
As one can see, the clover action suffers from serious deviations from 
$I\neq0$, especially for hopping parameters in the $b$-quark region.
On the other hand, the OK action's inconsistency is
statistically consistent with~$I=0$ up to the largest masses considered.

Another way to see the effects of improvement is to look at the
hyperfine splittings.
Let us define
\begin{eqnarray}
    \Delta_1 & = & M_1^{\mathrm{V}}-M_1^{\mathrm{PS}}, \\
    \Delta_2 & = & M_2^{\mathrm{V}}-M_2^{\mathrm{PS}},
\end{eqnarray}
where V and PS denote vector and pseudoscalar mesons.
The rest-mass splitting $\Delta_1$ is 
accurate at the tree level, thanks to the clover term, but the 
kinetic-mass splitting $\Delta_2$
has contributions from higher-dimension corrections such as 
$\{\bm{\gamma}\cdot\bm{D}, i\bm{\Sigma}\cdot\bm{B}\}$, which are 
improved (unimproved) with the OK (clover) action.

\begin{figure}[btp]
\begin{center}
\includegraphics[width=.6\textwidth]{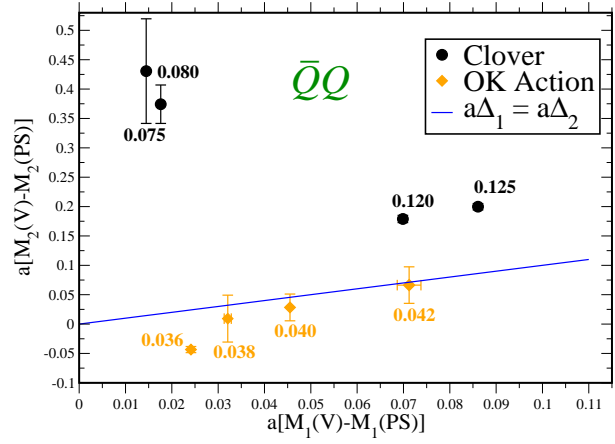}
\caption{Kinetic-mass hyperfine splitting \emph{vs}.\ rest-mass
hyperfine splitting for the heavy-heavy system.}
\label{fig:fig2}
\end{center}
\end{figure}

\begin{figure}[btp]
\begin{center}
\includegraphics[width=.6\textwidth]{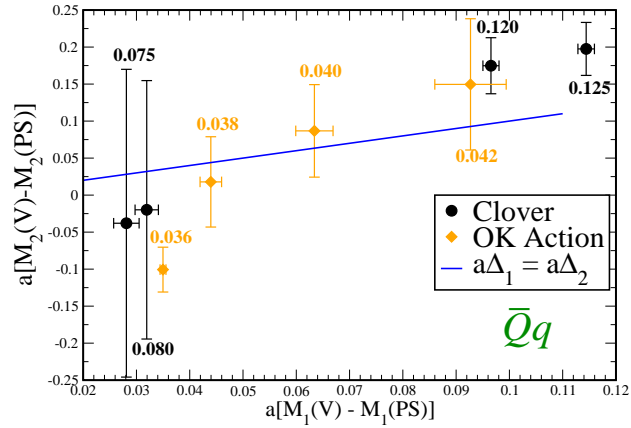}
\caption{Kinetic-mass hyperfine splitting \emph{vs}.\ rest-mass
hyperfine splitting for the heavy-light system.}
\label{fig:fig3}
\end{center}
\end{figure}

In Fig.~\ref{fig:fig2} we plot $a\Delta_2$ vs.\ $a\Delta_1$ for 
quarkonium at the same values of $\tilde{\kappa}_t$ as above.
Ideally, the data would land on the line $a\Delta_2=a\Delta_1$.
We see that the OK data fare much better than the clover data.
In Fig.~\ref{fig:fig3} we plot $a\Delta_2$ vs.\ $a\Delta_1$ for a
heavy-light meson.
In this case, the clover action fares well to begin with, and our 
statistical errors are not small enough to test whether the OK action 
is an improvement.

\section{Outlook}
\label{sec:outlook}

Our preliminary analysis of the OK action is encouraging. It
shows clear improvements in the inconsistency $I$ and the kinetic-mass
hyperfine splittings in the heavy-heavy system. For the
hyperfine splitting in the heavy-light system, the improvement is not 
yet clear, because the clover action already works well, so a decisive 
test requires higher statistics.
% The present code for inverting the OK Dirac operator, not yet
% optimized and without full tadpole improvement, runs 100 times slower than
% the clover inverter.
% Further study is underway. 
The code for the OK inverter is still under development.
The next step is to finish coding of the $c_5$ term
with the correct $u_0$ factors, to be followed by a thorough 
optimization.
Thereafter, we plan on using the OK action for 
charm and bottom physics.

Computations for this work were carried out in part on facilities of
the USQCD Collaboration, which are funded by the Office of Science of
the United States Department of Energy.
This work was supported in part by the U.S. Department of Energy under 
Grants No.~DE-FC02-06ER41446 (C.D., M.B.O), by
the National Science Foundation under Grants No.~PHY-0555243, No.~PHY-0757333,
No.~PHY-0703296 (C.D., M.B.O),
and by Universities Research Association, Inc.~(M.B.O.).
Fermilab is operated by Fermi Research Alliance, LLC, under Contract
No.~DE-AC02-07CH11359 with the U.S. Department of Energy.

\end{document}